\documentstyle[prb,aps,floats,epsf]{revtex}

\draft

\title{Ground-State Phase Diagram
of Quantum Heisenberg Antiferromagnets \\
on the Anisotropic Dimerized Square Lattice}

\author{Munehisa Matsumoto,$^1$ Chitoshi Yasuda,$^1$
Synge Todo,$^{1,2}$ and Hajime Takayama$^1$}

\address{$^1$Institute for Solid State Physics, University of Tokyo,
Kashiwa 277-8581, Japan}

\address{$^2$Theoretische Physik, Eidgen\"ossische Technische
Hochschule, CH-8093 Z\"urich, Switzerland}

\date{\today}

\begin{document}

\twocolumn[\hsize\textwidth\columnwidth\hsize\csname @twocolumnfalse\endcsname
\maketitle

\begin{abstract}
The $S=1/2$ and $S=1$ two-dimensional quantum Heisenberg antiferromagnets on
the anisotropic dimerized square lattice are investigated by the quantum
Monte Carlo method.  By finite-size-scaling analyses on the correlation
lengths, the ground-state phase diagram parametrized by strengths of the
dimerization and of the spatial anisotropy is determined much more
accurately than the previous works.  It is confirmed that the quantum
critical phenomena on the phase boundaries belong to the same
universality class as that of the classical three-dimensional Heisenberg
model.  Furthermore, for $S=1$, we show that all the spin-gapped phases,
such as the Haldane and dimer phases, are adiabatically connected in the
extended parameter space, though they are classified into different
classes in terms of the string order parameter in the one-dimensional,
i.e., the zero-interchain-coupling, case.
\end{abstract}


\vspace*{1.0em}

]\narrowtext

\section{Introduction}

Recently many low-dimensional antiferromagnets with excitation modes
separated from a ground state by a finite energy gap have been
synthesized, and effects of impurities or magnetic fields on those
materials have been investigated experimentally in relation to the
impurity-induced long-range order (LRO) and magnetic-field-induced
LRO. For example, there are the $S=1/2$ quasi-one-dimensional (Q1D)
Heisenberg antiferromagnet (HAF) with bond dimerization,
CuGeO$_3$,~\cite{hase} $S=1$ Q1D HAF's, NENP,~\cite{meyer}
NDMAZ,~\cite{honda1} NDMAP,~\cite{honda2} and
PbNi$_2$V$_2$O$_8$,~\cite{uchiyama} and $S=1$ Q1D HAF's with bond
alternation, NTEAP~\cite{hagiwara} and NTENP.~\cite{narumi1} Those
materials have attracted our interest since they reveal various aspects
of the quantum phase transition between the quantum-disordered
spin-gapped phase and the classical (N\'eel) long-range-ordered phase.

Intrachain spin interaction with or without bond alternation and
interchain interaction are considered to be the most basic ingredients
to understand the quantum phase transitions mentioned above.  More
explicitly, they are expected to be modeled effectively by the quantum
HAF on the anisotropic dimerized square lattice, which is described by
the Hamiltonian:
\begin{eqnarray}
   \label{ham}
   {\cal H}&=&\sum_{i,j}{\bf S}_{2i,j}\cdot{\bf S}_{2i+1,j}
    +\alpha\sum_{i,j}{\bf S}_{2i+1,j}\cdot{\bf S}_{2i+2,j} \\
    &+&J'\sum_{i,j}{\bf S}_{i,j}\cdot{\bf S}_{i,j+1} \ . \nonumber
\end{eqnarray}
Here ${\bf S}_{i,j}$ is the quantum spin operator at site $(i,j)$ on the
square lattice. The first two terms in r.h.s.\ represent the
one-dimensional antiferromagnetic (AF) Heisenberg chains with
alternating coupling constants, 1 and $\alpha$ $(0\leq \alpha \leq 1)$,
and the last term does the AF interchain exchange interaction ($J' \geq
0$).  We choose the $x$-axis as being along the chain direction and the
$y$-axis as in the perpendicular one.  The bond arrangement of this
model is shown in Fig.~\ref{notation}.

The ground state of decoupled chains, i.e., $J'=0$, has been well
established.  In particular, that of the $S=1/2$ chain~\cite{bulaevskii}
is the dimer state with a finite spin gap except for the uniform case
($\alpha=1$), which has a critical ground state.  On the other hand, for
the $S=1$ chain there exist two spin-gapped phases: the Haldane
phase~\cite{haldane} at $\alpha > \alpha_{\rm c}$ and the dimer phase at
$\alpha < \alpha_{\rm c}$.\cite{singh} At the critical point
$\alpha=\alpha_{\rm c}$ between these two phases the gap vanishes. The
value of $\alpha_{\rm c}$ has been estimated to be
0.5879(6).\cite{singh,kohno} In both cases, the critical point is
considered to belong to the Gaussian universality class.\cite{singh}

For the AF LRO to appear the higher-dimensionality effect, i.e., the
interchain interaction $J'$, is indispensable.  In most of the numerical
works reported so far, the effect of interchain coupling has been
examined in certain approximated or perturbed ways.  For example, Sakai
and Takahashi~\cite{sakai} estimated the critical strength, $J'_{\rm
c}$, for the uniform case ($\alpha=1$) by the exact diagonalization
method for the intrachain interactions combined with the mean-field
approximation for the interchain interaction, and obtained
$J'^{(S=1/2)}_{\rm c}=0$ and $J'^{(S=1)}_{\rm c}\geq 0.025$.  More
recently, Koga and Kawakami~\cite{koga} investigated the $S=1$ model by
the cluster-expansion method, and obtained $J'_{\rm c} = 0.056(1)$ for
$\alpha = 1$.  However, there have been only a very limited number of
numerical works, in which both of the interchain and intrachain
interactions are treated on an equal footing.~\cite{katoh,kim} Such
numerical analyses are certainly required, since the mean-field-like
approximation is not necessarily appropriate even in the Q1D
regime.~\cite{affleck}

In the present paper, we report the results of quantum Monte Carlo (QMC)
simulations by using the continuous-imaginary-time loop
algorithm~\cite{evertz1,evertz2,harada,todo2} on the $S=1/2$ and $S=1$
HAF model described by Eq.~(\ref{ham}).  The present paper is organized
as follows.  In Sec.~\ref{method-section}, the method of our numerical
analyses is explained.  In Secs.~\ref{result-section-spin-half} and
\ref{result-section-spin-one}, the ground-state phase diagram
parameterized by the strength of the bond alternation, $\alpha$, and
that of the interchain coupling, $J'$, is determined precisely for
$S=1/2$ and $S=1$, respectively.  Especially, for the $S=1$ system with
$\alpha=1$, we obtain $J'_{\rm c} = 0.043648(8)$, which is consistent
with $0.040(5)$ suggested by the recent QMC work,\cite{kim} but is much
more accurate.  Furthermore, both in the $S=1/2$ and $S=1$ systems, the
quantum phase transitions between the spin-gapped phases and the AF-LRO
phase are confirmed to belong to the same universality class with that
of the 3D classical Heisenberg model: the exponent of the correlation
lengths is $\nu=0.71(3)$ for $S=1/2$ and $\nu=0.70(1)$ for $S=1$, which
coincides fairly well with that of the latter model,
$\nu=0.7048(30)$.~\cite{chen} We also show the results on the
correlation length and the gap in the spin-gapped phase.  In
Sec.~\ref{discussions-section}, the topology of the phase diagram is
discussed in detail based on the result of the present QMC calculation.
We show that all the spin-gapped phases, such as the Haldane and dimer
phases, are adiabatically connected with each other in the extended
phase parameter space.  This is in a sharp constrast to the strict 1D
case, in which the spin-gapped phases are classified into different
classes in terms of the so-called string-order parameter.\cite{denNijs}
It is of interest that the 1D spin-gap phases, which have different
hidden symmetry, are connected without encountering any singularity in
the 2D phase diagram.  The final section is devoted to the concluding
remarks.

\begin{figure}[tbp]
 \centerline{\epsfxsize=0.22\textwidth\epsfbox{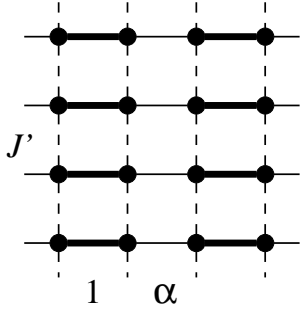}}
 \vspace*{1em}
 \caption{Anisotropic dimerized square lattice with alternating
 intrachain coupling of strength 1 (thick solid line) and $\alpha$
 (solid line), and the interchain coupling of strength $J'$ (dashed
 line).}  \label{notation}
\end{figure}

\section{Method}
\label{method-section}

We consider the system descrived by the Hamiltonian~(\ref{ham}) with
$S=1/2$ and $S=1$.  The real-space size is $L_x \times L_y$ and the inverse
temperature, i.e., the imaginary-time size, is $\beta=1/T$.  Periodic
boundary conditions are imposed in the $x$- and $y$-directions.  We use
the continuous-imaginary-time loop algorithm with multi-cluster
update.\cite{evertz1,evertz2} Especially, for the $S=1$ system we adopt
the subspin-representation technique,\cite{harada,todo2} in which the
$S=1$ system is represented by an $S=1/2$ system with special boundary
conditions in the imaginary-time direction.  By using these techniques,
we can perform the simulatation up to $L_x \times L_y = 336 \times 48$
with $\beta = 100$ for the $S=1$ case without encountering any
difficulty.

The imaginary-time dynamical structure factor is defined by
\begin{eqnarray}
&& S_{\rm d}(q_{x},q_{y},\omega) \\
&& \mbox{} = 
\frac{1}{L_{x}L_{y}\beta}
\sum_{i,j}\int_{0}^{\beta} \!\! dt \, dt'\,
{\rm e}^{- i {\bf q}\cdot({\bf r}_{i}-{\bf r}_{j})- i \omega(t-t')}
\langle S_{i}^{z}(t)S_{j}^{z}(t') \rangle, \nonumber
\end{eqnarray}
where ${\bf q}=(q_{x},q_{y})$ is the wave-number vector, $S_{i}^{z}(t)$
is the $z$-component of the spin on site $i$ at imaginary time $\tau$,
and $\langle\cdots\rangle$ is the thermal average.  By using
$S_{\rm d}(q_x,q_y,\omega)$, the staggered correlation length along the
$x$-direction, $\xi_x$, is then evaluated by the second-moment
method,~\cite{todo2,cooper}
\begin{equation}
 \xi_x = {L_x \over 2\pi}\sqrt{ {S_{\rm d}(\pi,\pi,0) \over 
S_{\rm d}(\pi+2\pi/L_{x},\pi,0) } -1}.
\end{equation}
The correlation length in the $y$-direction, $\xi_y$, and that in the
imaginary-time ($\tau$) direction, $\xi_\tau$, which is related to the gap $\Delta$ by
$\Delta = 1/\xi_\tau$, are calculated similarly.  Finally the staggered
susceptibility, $\chi_{\rm s}$, is evaluated by
\begin{eqnarray}
\chi_{\rm s} &=&S_{\rm d}(\pi,\pi,0) \\
\mbox{} &=& 
\frac{1}{L_{x}L_{y}\beta}
\sum_{i,j}\int_{0}^{\beta} \!\! dt \, dt'\,
e^{-i {\bf \pi} \cdot ({\bf r}_i - {\bf r}_j)}
\langle S_{i}^{z}(t)S_{j}^{z}(t') \rangle. \nonumber
\end{eqnarray}
All the structure factors are calculated by using the improved
estimators.\cite{baker} The period of $10^2$--$10^3$ Monte Carlo
steps (MCS) is used for thermalization and that of $10^3$--$10^5$
MCS for the evaluation of physical quantities.

Near the critical point $(\alpha_{\rm c},J'_{\rm c})$ of the ground-state 
transition, the correlation lengths diverge as
\begin{eqnarray}
\xi_x, \xi_y\sim t^{-\nu}\\
\xi_{\tau}\sim t^{-z\nu}= t^{-\nu},
\end{eqnarray}
where $t$ is the distance from the critical point and $\nu$ is the
critical exponent for the correlation length. Here we have put $z=1$
assuming the Lorenz invariance.\cite{chakravarty} Furthermore, the
following finite-size-scaling (FSS) formula\cite{barber} holds near
$(\alpha_{\rm c},J'_{\rm c})$ and $T=0$ for systems with the fixed ratio
$L_x:L_y:\beta$,
\begin{equation}
\xi_x/L_x \simeq f(t L_x^{1/\nu}, L_x^{z}T) =f(t L_x^{1/\nu}), 
\label{fss-formula-for-xi}
\end{equation}
and similar ones for $\xi_y$ and $\xi_{\tau}$, and
\begin{equation}
\chi_{\rm s}\simeq L_x^{\gamma/\nu}g(t L_x^{1/\nu},L_x^{z}T)=
L_x^{\gamma/\nu}g(t L_x^{1/\nu}).
\label{fss-formula-for-chi-s}
\end{equation}
Here $f$ and $g$ are scaling functions and $\gamma$ the exponent for
$\chi_{\rm s}\ (\sim t^{-\gamma})$. Note that $L_xT$ is put constant in
the above equations.  We assume a polynomial up to the second order for
the scaling functions. By using least-squares fitting, we obtain the
critical point $(\alpha_{\rm c}, J'_{\rm c})$ and the associated
critical exponents $\nu$ and $\gamma$.

In addition, at some points in the spin-gapped phase, we explicitly
evaluate the correlation lengths, $\xi_x$ and $\xi_y$, and the gap,
$\Delta$, at $T=0$ in the thermodynamic limit $L_x, L_y \rightarrow
\infty$.  For this purpose we extrapolate the simulated data first to
the ground state $T\rightarrow 0$ and then to the thermodynamic limit
$L_x, L_y \rightarrow \infty$.

\section{Results for $\bf S=1/2$}
\label{result-section-spin-half}

\begin{figure}[tbp]
 \centerline{\epsfxsize=0.45\textwidth\epsfbox{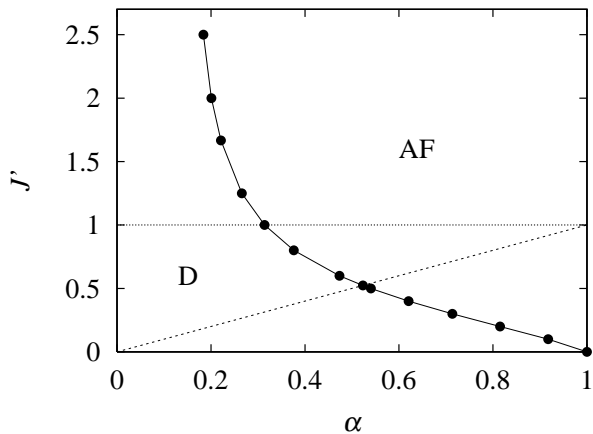}}
 \vspace*{.5em}
 \caption{The $\alpha$-$J'$ phase diagram of the $S=1/2$ system at zero
 temperature. The (spin-gapped) dimer phase and the AF-LRO phase are
 described by D and AF, respectively.  The statistical error of each
 data point is smaller than the symbol size.  The dotted and dashed
 lines denote $J'=1$ and $\alpha=J'$, respectively.}
 \label{gs_phase_diagram_spin_half}
\end{figure}

\subsection{Ground-state phase diagram}

\begin{figure}[tbp]
 \centerline{\epsfxsize=0.45\textwidth\epsfbox{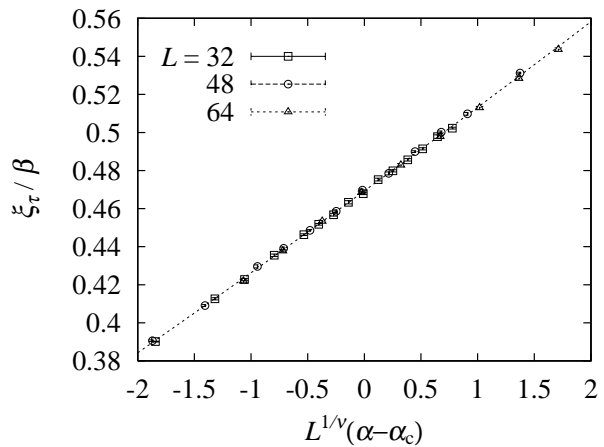}}
 \vspace*{.5em}
 \caption{Finite-size scaling plot of the inverse gap, $\xi_\tau$, for
 $S=1/2$ with $J'=1$ and $L_x:L_y:\beta = 1:1:1$.  The critical coupling
 $\alpha_{\rm c}$ and the exponent $\nu$ are estimated as to be
 0.31407(5) and 0.71(1), respectively.  The dashed line represents the
 scaling function, which is approximated by a polynomial of order $2$.}
 \label{fss_spin_half}
\end{figure}

In Fig.~\ref{gs_phase_diagram_spin_half} we show the ground-state phase
diagram of the $S=1/2$ system obtained by the FSS analysis explained in
the previous section.  As an example of the FSS analysis, we show in
Fig.~\ref{fss_spin_half} that of $\xi_\tau$ against $\alpha$ for $J'=1$
(dotted line in Fig.~\ref{gs_phase_diagram_spin_half}).  The aspect
ratio of the $(2+1)$-dimensional system is taken as $L_x:L_y:\beta =
1:1:1$.  By the least-squares fitting, the exponent $\nu$ and the
critical coupling $\alpha_{\rm c}$ are estimated as 0.71(1) and
0.31407(5), respectively. Here, the figure in parentheses denotes the
statistical error ($1\sigma$) in the last digit.  We also perform the
same analyses on other lines in the $\alpha$-$J'$ plane, whose results
are presented by the solid circles in
Fig.~\ref{gs_phase_diagram_spin_half}.  For example, on the line $\alpha
= J'$ (dashed line in Fig.~\ref{gs_phase_diagram_spin_half}), we obtain
$\alpha_{\rm c} = J'_{\rm c} = 0.52337(3)$ and $\nu = 0.71(3)$.  In the
phase diagram we can see that the ground state of the chain ($J'=0$) is
the dimer state with a spin gap except for
$\alpha=1$,~\cite{sakai,aoki,affleck2,sandvik} and that the region of
the AF-LRO phase enlarges monotonically as $J'$ increases.

Our phase diagram is qualitatively the same as that of Katoh and Imada
(KI),~\cite{katoh} but not quantitatively.  In particular, we obtain the
critical point, $\alpha_{\rm c}=0.31407(5)$, on the line
$J'=1$ (dotted line in Fig.~\ref{gs_phase_diagram_spin_half}), which is
significantly smaller than their estimate $\alpha_{\rm c} = 0.398$.
More importantly, in the present simulation, the critical exponent $\nu$
on the transition points is evaluated as $\nu = 0.71(1)$, which is
consistent with $\nu = 0.7048(30)$ for the 3D classical Heisenberg
model.~\cite{chen} The similar results have been obtained for the 2D
$1/5$-depleted HAF model.\cite{troyer}  On the other hand, KI concluded
$\nu=1$.  The reason of the these discrepancies might be due to the
smallness of the system sizes and the inverse temperature used in the
study by KI.

\subsection{Correlation lengths and the gap}

\begin{figure}[tbp]
 \centerline{\epsfxsize=0.43\textwidth\epsfbox{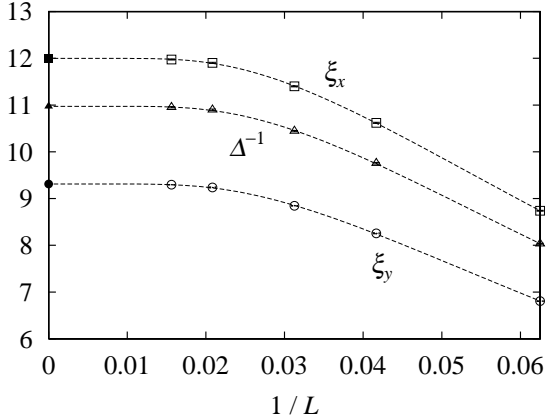}}
 \vspace*{.5em}
 \caption{System-size dependence of the correlation lengths and the
 inverse gap at zero temperature for $S=1/2$ with $\alpha=J'=0.5$. The
 extrapolated values are denoted by solid symbols.}
 \label{fig_corr_one_half}
\end{figure}

We also evaluate explicitly the ground-state correlation lengths and the
gap on some points in the dimer phase by using the dynamic structure
factors. Unless the points are very close to the critical line, these
quantities in each systems with $L$ ($=L_x=L_y$) saturate to the
ground-state values at temperatures we have simulated. For example, the
$T$-dependences of these quantities are not to be discernible at
$T=0.05$ and $0.01$ for $\alpha=J'=0.4$ and 0.5, respectively.  On the
other hand, the $L$-dependence still remains in sizes we have
calculated. The $L$-dependences of the ground-state spatial correlation
lengths and the gap are shown in Fig.~\ref{fig_corr_one_half} for
$\alpha=J'=0.5$, which is close to the critical point $\alpha_{\rm
c}=J'_{\rm c}=0.52337(3)$.  Their values in the thermodynamic
limit are estimated by fitting $\xi_k(L)$ to $\xi_k(L)= \xi-b
\exp (-c L)$, where $k=x$, $y$, or $\tau$, $\xi$ is the value in the thermodynamic limit, and
$b$ and $c$ are fitting parameters. As a result, we obtain
$\xi_{x}=3.0089(9)$, $\xi_{y}=2.2097(6)$ and $\Delta=0.32261(4)$ for
$\alpha=J'=0.4$ and $\xi_{x}=11.998(9)$, $\xi_{y}=9.312(10)$, and
$\Delta=0.0913(2)$ for $\alpha=J'=0.5$. As $\alpha$ ($=J'$) becomes
smaller, i.e., the system becomes more distant from the critical point,
$\Delta$ becomes larger and $\xi$ smaller.

\begin{figure}[t]
 \centerline{\epsfysize=6cm\epsfbox{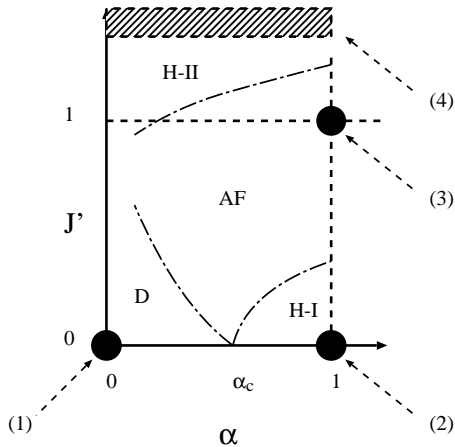}}
 \vspace*{.5em}
 \caption{Schematic ground-state phase diagram of the S=1 system
 obtained before the present QMC analysis.  Filled circles are the
 points where the corresponding ground state is well-established: (1)
 the dimer phase (D), (2) $x$-parallel Haldane phase (H-I), (3)
 the AF-LRO phase (AF), and (4) $y$-parallel Haldane phases (H-II).}
 \label{schematic-phase-diagram}
\end{figure}

\section{Results for $\bf S=1$}
\label{result-section-spin-one}

\subsection{Overview on the phase diagram}

Before going into the QMC analysis, let us here summarize the
ground-state phase diagram of the $S=1$ system argued so far, which is
shown in Fig.~\ref{schematic-phase-diagram}.  For some points in the
phase diagram the ground state is well understood by the previous
theoretical and numerical studies: (1) $(\alpha,J')=(0,0)$: The system
consists of a set of the isolated antiferromagnetically-coupled spin
pairs.  The ground state is a trivial tensor product of dimer singlets
sitting on each bond.  (2) $(\alpha,J')=(1,0)$: The system consists of
isolated $x$-parallel Haldane chains.  (3) $(\alpha,J')=(1,1)$: The
system is a uniform and isotropic 2D HAF.  There exists an AF LRO in the
ground state.\cite{kubo} (4) $J'=\infty$: The system consists of
$y$-parallel Haldane chains.  Note that in this limit the value of
$\alpha$ becomes irrelevant (see also discussions in
Sec.~\ref{discussions-section}).

In their analysis by the cluster expansion method, Koga and
Kawakami\cite{koga} derived three phases in the Q1D region, and they
called the regions which includes point (1), (2) and (3) the dimer
phase, the Haldane phase, and the AF-LRO phase, respectively. The region
near point (4) is another Haldane phase.  Therefore we call here the
region which includes point (2) the Haldane I (H-I) phase and the one
which includes the line (4) the Haldane II (H-II) phase. For the uniform
systems with $\alpha=1$ the H-I and H-II phases are equivalent when we
exchange the roles of the $x$-axis and the $y$-axis, and of $J'$ and
$1/J'$.

\begin{figure}[tbh]
 \hspace*{2em}(a) \\
 \vspace*{-1.5em}
 \centerline{\epsfysize=4.7cm\epsfbox{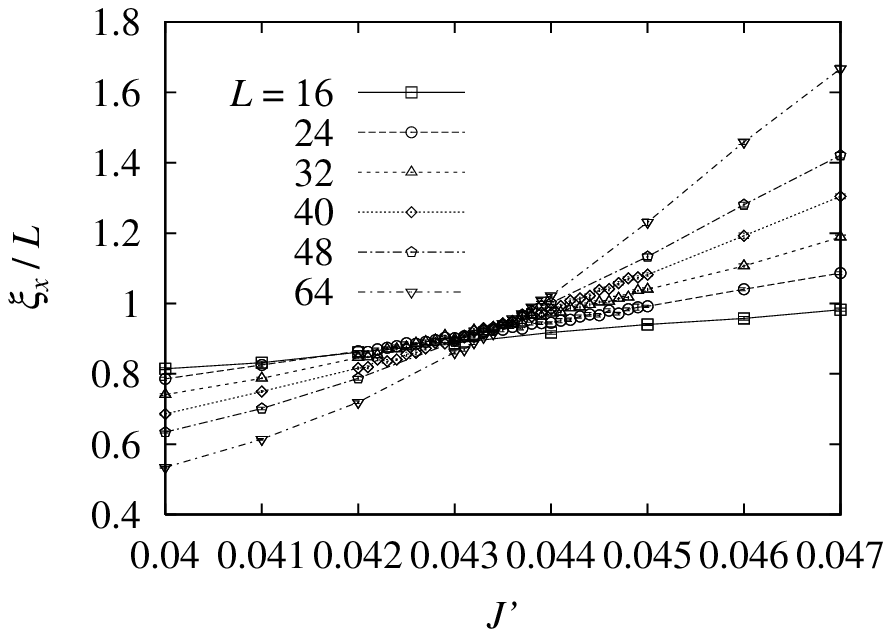}}

 \hspace*{2em}(b)\\
 \vspace*{-1.5em}
 \centerline{\epsfysize=4.7cm\epsfbox{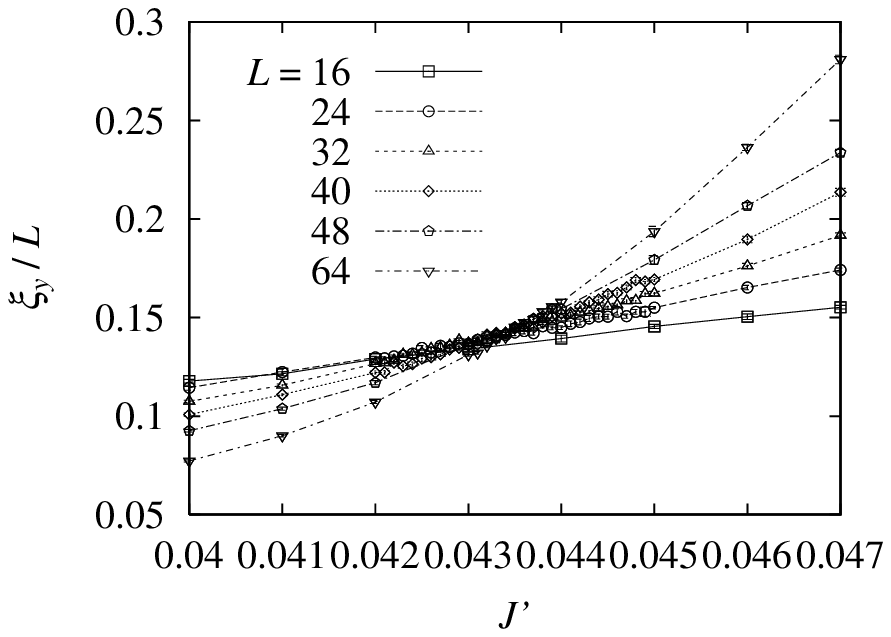}}

 \hspace*{2em}(c)\\
 \vspace*{-1.5em}
 \centerline{\epsfysize=4.7cm\epsfbox{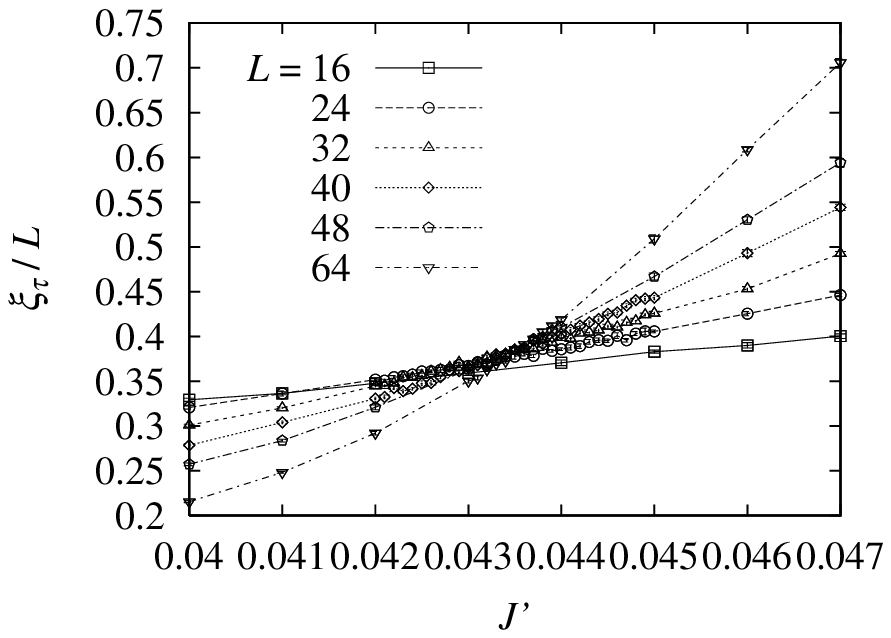}}
 \vspace*{.0em}
 \caption{Plot of the correlation lengths, (a) $\xi_{x}/L_{x}$, (b)
 $\xi_{y}/L_{y}$, and (c) $\xi_{\tau}/\beta$, as a function of $J'$ for
 $S=1$ with $\alpha=1$ and $L_x:L_y:\beta=1:1:1$.}  \label{raw_data}
\end{figure}

\begin{figure}[tb]
 \hspace*{2em}(a) \\
 \vspace*{-1.5em}
 \centerline{\epsfysize=4.7cm\epsfbox{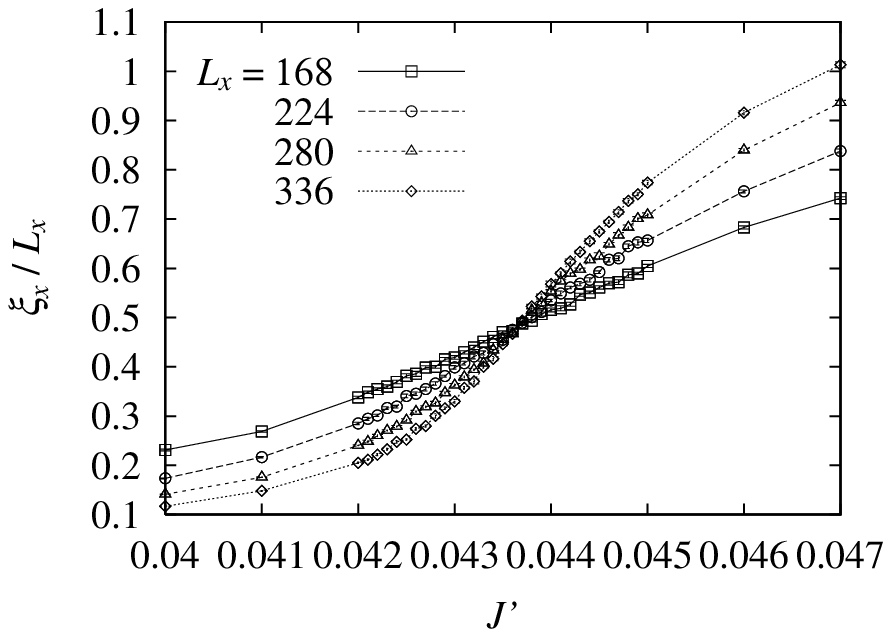}}
 \hspace*{2em}(b) \\
 \vspace*{-1.5em}
 \centerline{\epsfysize=4.7cm\epsfbox{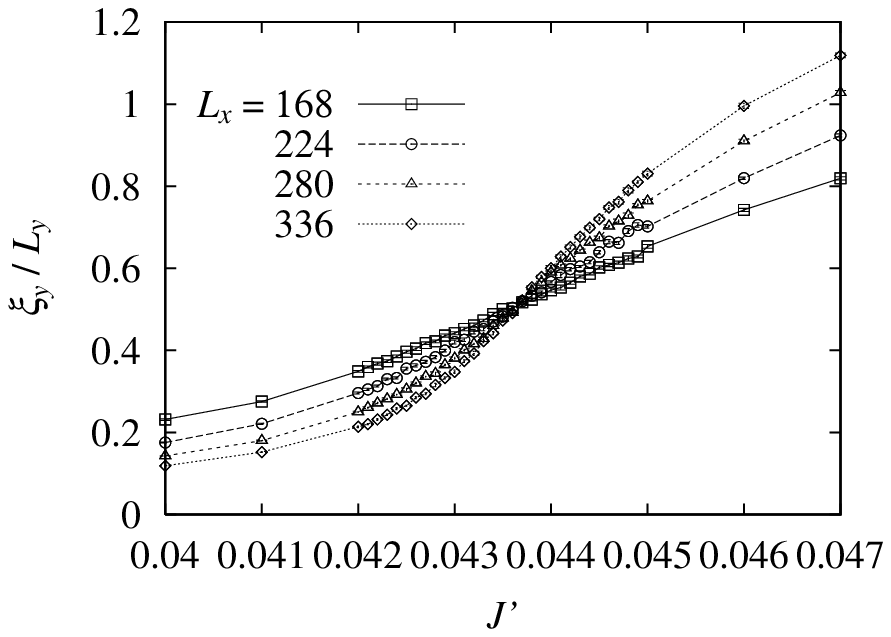}}
 \hspace*{2em}(c) \\
 \vspace*{-1.5em}
 \centerline{\epsfysize=4.7cm\epsfbox{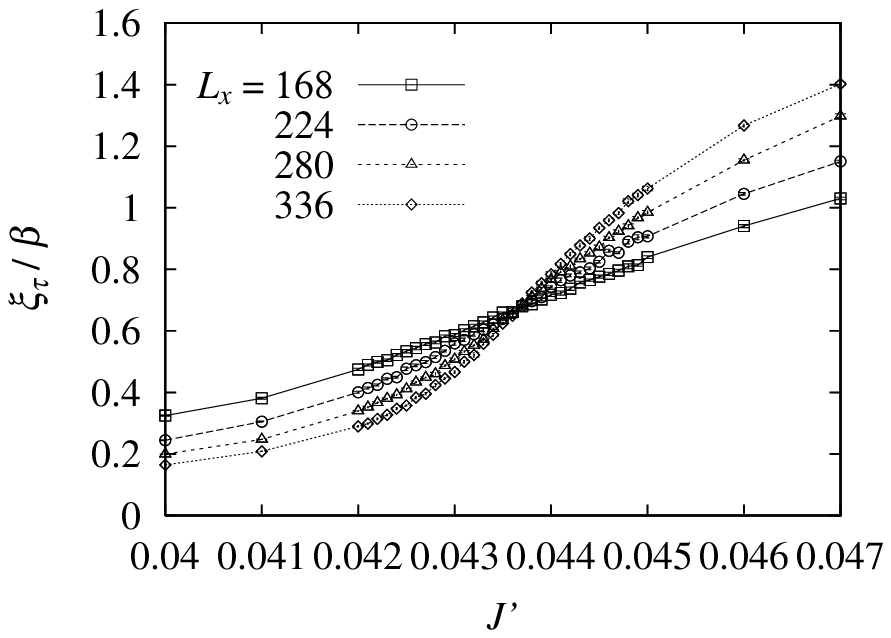}}
 \vspace*{.0em}
 \caption{Plot of the correlation lengths, (a) $\xi_{x}/L_{x}$, (b)
 $\xi_{y}/L_{y}$, and (c) $\xi_{\tau}/\beta$, as a function of $J'$ for
 $S=1$ with $\alpha=1$ and $L_{x}:L_{y}:\beta=7:1:2$.}
 \label{raw_data_large_system}
\end{figure}

\subsection{Haldane-AF phase transition in the non-dimerized system}

To demonstrate our FSS analyses, let us begin with critical behavior
near $(1,J'_{\rm c})$, which separates the H-I and AF phases ($J'_{\rm c}
\simeq 0.04$ due to Ref.~\onlinecite{kim}). We sweep $J'$ near supposed
$J'_{\rm c}$ with $\alpha$ fixed to unity.  In Fig.~\ref{raw_data},
$\xi_x/L_x$, $\xi_y/L_y$, and $\xi_{\tau}/\beta$ of the systems with
$L_x=L_y=\beta\equiv L$ ($=16$, 24,$\cdots$, 64) are plotted.  As one
sees immediately, the data suffer from quite large corrections to
scaling, i.e., the crossing point of the scaled correlation lengths with
two different $L$'s clearly shifts to larger $J'$ as the system size
increases.  We attribute these large corrections to the strong spatial
anisotropy in the coupling constants ($J' \ll 1$).  Indeed, the value of
$\xi_y$ at $J'=0.0435$ is $9.29(7)$ (see Fig.~\ref{raw_data}~(b)) even
for the largest system size $L=64$, which is quite smaller than those
in the other directions ($\xi_x=60.4(3)$ and $\xi_{\tau}=24.7(2)$).
Among the three correlation lengths, $\xi_x$ is the largest: the growth
of the correlation is dominated only by the system size in the
$x$-direction.  This indicates that we need larger lattices, especially
in the $x$-direction, in order to perform a precise FSS analysis.

\begin{figure}[tbp]
 \centerline{\epsfysize=5cm\epsfbox{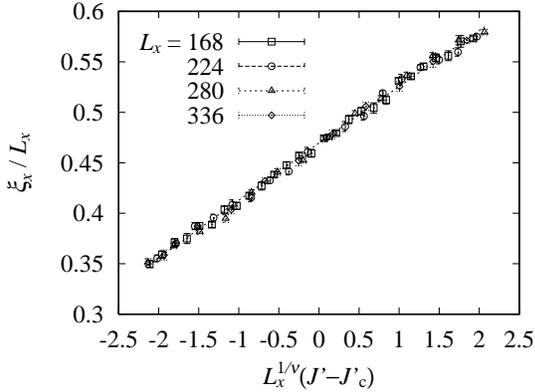}}
 \vspace*{.5em}
 \caption{Finite-size-scaling plot of the correlation length, $\xi_{x}$
 for $S=1$ with $\alpha=1$ and $L_{x}:L_{y}:\beta=7:1:2$.  The critical
 coupling $J'_{\rm c}$ and the exponent $\nu$ are estimated as to be
 0.043648(9) and 0.69(1), respectively, by the least-squares fitting.
 The dashed line represents the scaling function, which is approximated
 by a polynomial of order $2$.}  \label{fss_plot_corrx_large_system}
\end{figure}

To simulate larger lattices with minimal costs, we therefore optimize
the aspect ratio, $L_x:L_y:\beta$, as explained below.  Expecting that
the scaled correlation lengths, $\xi_x/L_x$, $\xi_y/L_y$, and
$\xi_\tau/\beta$, become nearly equal with each other at the critical
point, we set the aspect ratio $L_x:L_y:\beta$ as $7:1:2$ based on the
data presented in Fig.~\ref{raw_data}. With this ratio we simulate
systems with $L_x=168$, 224, 280, and 336 and perform the FSS analyses.
The raw data of $\xi_x/L_x$, $\xi_y/L_y$, and $\xi_{\tau}/\beta$ with
this aspect ratio are shown in Fig.~\ref{raw_data_large_system}.  Now
$\xi_y=12.02(5)$ even for the smallest system size ($L_x=168$ and
$L_y=24$) at $J'=0.0435$.  The ratios $\xi_x/L_x$, $\xi_y/L_y$ and
$\xi_\tau/\beta$ in a common range of $J'$ become nearly equal, and
corrections to scaling become much smaller than in the previous ones as
we expected.  The FSS plot for $\xi_{x}$ is shown in
Fig.~\ref{fss_plot_corrx_large_system}.  The resultant $J'_{\rm c}$ and
$\nu$ are as follows: $(J'_{\rm c},\nu)=(0.043648(9),0.69(1))$ from
$\xi_x$, $(0.043649(8),0.71(1))$ from $\xi_y$, and
$(0.043648(7),0.69(1))$ from $\xi_{\tau}$.
Averaging these three values we conclude with
\begin{eqnarray}
J'_{\rm c}& = & 0.043648(8)\label{precise_Jc}
\end{eqnarray}
and
\begin{eqnarray}
\nu &=& 0.70(1).\label{nu_S1}
\end{eqnarray}

\begin{figure}[t]
 \centerline{\epsfysize=5cm\epsfbox{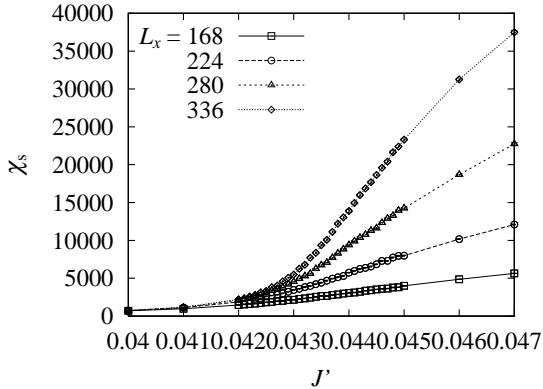}}
 \vspace*{.5em}
 \caption{Plot of the staggered susceptibility, $\chi_{\rm s}$, as a
 function of $J'$ for $S=1$ with $\alpha=1$ and
 $L_{x}:L_{y}:\beta=7:1:2$.}  \label{raw_data_ssus}
\end{figure}
\begin{figure}[t]
 \centerline{\epsfysize=5cm\epsfbox{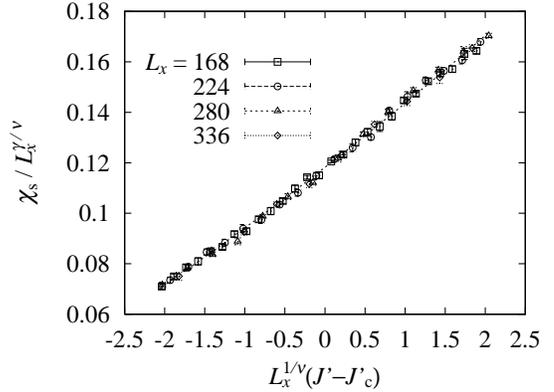}}
 \vspace*{.5em}
 \caption{Finite-size-scaling plot of the staggered susceptibility,
 $\chi_{\rm s}$, for $S=1$ with $\alpha=1$ and
 $L_{x}:L_{y}:\beta=7:1:2$.  The values of $J'_{\rm c}$ and $\nu$ are
 fixed to $0.043648$ and $0.70$, respectively.  The exponent $\gamma$ is
 estimated as to be 1.373(3) by the least-squares fitting. The dashed
 line represents the scaling function, which is approximated by a
 polynomial of order $2$.}  \label{fss_plot_ssus}
\end{figure}

Fixing the value of $J'_{\rm c}$ and $\nu$ thus determined, we next
perform the FSS analysis on the staggered susceptibility $\chi_{\rm
s}$. The raw data of $\chi_{\rm s}$ vs. $J'$ are shown in
Fig.~\ref{raw_data_ssus} and the fitting result is shown in
Fig.~\ref{fss_plot_ssus}. The latter yields
\begin{equation}
\gamma=1.373(3).
\label{gamma_S1}
\end{equation}
The exponents $\nu$ and $\gamma$ we have obtained for the $S=1$ system
again agree with those of the 3D classical Heisenberg model.~\cite{chen}

The critical point obtained just above is consistent with the previous
result by the method involving the mean-field approximation, $J'_{\rm
c}\geq 0.025$,\cite{sakai} and also with that by the recent QMC method,
$J'_{\rm c}=0.040(5)$.\cite{kim} On the other hand, the present result
significantly differs from that obtained by the cluster-expansion
method, $J'_{\rm c} = 0.56(1)$.\cite{koga} The present FSS analyses
on our extensive QMC results make us possible
to obtain the critical value with much
higher accuracy than the other methods.

\subsection{Ground-state phase diagram}

\begin{figure}[t]
 \centerline{\epsfysize=6.5cm\epsfbox{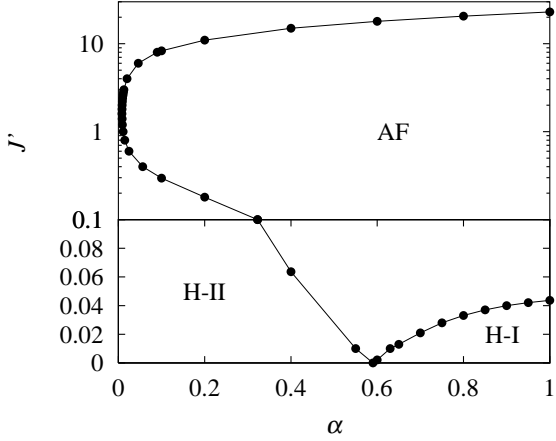}}
 \vspace*{.5em}
 \caption{Ground-state phase diagram of the $S=1$ system.  The $J'$ axis
 is logarithmically scaled for $J'>0.1$ for convenience.  The
 statistical error of each data point is smaller than the symbol size.}
 \label{gs_phase_diagram_spin_one}
\end{figure}

In a similar way as described in the previous subsection, we obtain
other critical points on the $\alpha$-$J'$ phase diagram as shown in
Fig.~\ref{gs_phase_diagram_spin_one}.  First, the scaled correlation
lengths, $\xi_x/L_x$, $\xi_y/L_y$, and $\xi_{\tau}/\beta$, are calculated
up to $L_x = 64$ with either $\alpha$ or $J'$ fixed.  Sweeping $J'$ or
$\alpha$ with sufficiently high resolution, we regard a crossing point
of these $\xi$'s as the critical point.  Note that the optimal aspect
ratio depends strongly on the value of $\alpha$ and $J'$.  However, we
adopt $L_x:L_y:\beta=1:1:1$ for simplicity.  Although the results thus
obtained suffer from relatively larger systematic corrections than those
presented in the last subsection, the absolute magnitude of the
systematic error in the estimates should be still smaller enough than the
symbol size in Fig.~\ref{gs_phase_diagram_spin_one}.

For some critical points the FSS analysis as in the previous subsection
is also carried out.  We obtain the exponents $\nu$ and $\gamma$ which
are consistent with Eqs.~(\ref{nu_S1}) and (\ref{gamma_S1}),
respectively. This supports that the quantum critical phenomena in the
$S=1$ system also belong to the same universality class as that of
the 3D classical Heisenberg model.  An exception is the 1D critical
point located at $(\alpha_{\rm c},J_{\rm c}')=(0.5879(6),0),$\cite{kohno}
which separates the dimer phase from the Haldane phase. The apparent value of
$\nu$ starts to deviate from (\ref{nu_S1}) when $\alpha$ becomes closer to
$\alpha_{\rm c}$.  This is attributed to the crossover to the
critical phenomena belonging to the Gaussian universality
class.\cite{singh} We confirm that the AF-LRO phase exists between the
two spin-gap phases at least down to $J'=0.01$ at $\alpha=\alpha_{\rm
c}$.  Although in the present simulation it is quite difficult to prove
the existence of the AF-LRO phase at smaller $J'$, we believe that the
the point $(\alpha_{\rm c},0)$ is tricritical: the 1D critical point is
unstable against an infinitesimal interchain coupling and the AF LRO
immediately appears as the same as in the $S=1/2$ uniform
chain.~\cite{sakai,aoki,affleck2,sandvik}

\begin{figure}[t]
 \centerline{\epsfysize=5cm\epsfbox{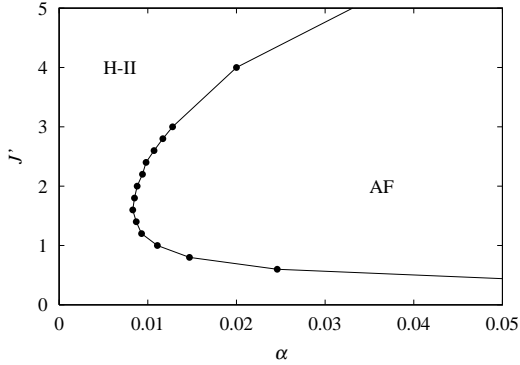}}
 \vspace*{.5em}
 \caption{Ground-state phase diagram of $S=1$ system in the
 strongly-dimerized region ($\alpha \le 0.05$).}
 \label{phase_diagram_near_the_connected_part}
\end{figure}

Interestingly, the H-II and D phases are adiabatically connected with
each other. The gapless AF-LRO phase does not touch the line of
$\alpha=0$ as seen in Fig.~\ref{phase_diagram_near_the_connected_part},
where the part of the whole phase diagram
(Fig.~\ref{gs_phase_diagram_spin_one}) near $\alpha=0$ is magnified.
Indeed, on the $\alpha=0$ line, which corresponds to the $S=1$ two-leg
ladder, it is shown that there exists no critical point by the recent
QMC study.\cite{ladder-paper} Thus, the D phase can be identified with
the H-II phase, and there are only two distinct spin-gap phases, H-I and
H-II in Fig.~\ref{gs_phase_diagram_spin_one}.  The closeness of the
critical line to the $\alpha=0$ line is due to the strong AF
fluctuations, which already exist in the two-leg ladder
system.\cite{ladder-paper}

\subsection{Correlation lengths and the gap}

We obtain the explicit values of $\xi_{x}$, $\xi_{y}$, and $\Delta$ in
the ground state at $(\alpha,J')=(1,0.04)$.  They are calculated for
systems with sizes $L_{x}=168$, 224, 280, and 336 and with the aspect
ratio $L_{x}:L_{y}=7:1$ at $T$ regarded as zero temperature.  Their
$T$-dependences are negligible at $T=0.01$.  We extrapolate the
finite-size data to the thermodynamic limit in the same way as explained
for the $S=1/2$ system.  We obtain $\xi_{x}=39.2(1)$, $\xi_{y}=5.67(1)$,
and $\Delta=0.0632(2)$.  As $J'$ becomes smaller, $\xi_{x}$ becomes
smaller and $\Delta$ larger to reach at $J'=0$ the single chain
values $\xi_{x}=6.0153(3)$ and $\Delta=0.41048(6)$.\cite{todo2}


\section{Discussions}
\label{discussions-section}

The analyses presented in the preceding section revealed that the
ground-state phase diagram of the $S=1$ system has a rather complicated
topology, i.e., the H-II and D phases are adiabatically connected with
each other, though the channel between them is quite narrow
(Fig.~\ref{gs_phase_diagram_spin_one}).  On the other hand, as
for the H-I and D phases, in the 1D system ($J'=0$) these two
spin-gapped phases are distinctively separated by the critical point at
$\alpha_{\rm c}=0.5879(6)$,\cite{singh,kohno} and they are distinguished
by the string-order parameter,\cite{denNijs} which is zero in the former
phase and finite in the latter one.  The transition can be viewed as a
rearrangement of dimer-singlet pattern between the (1,1)- and
(2,0)-valence-bond-solid (VBS) states.\cite{AKLT,arovas} We emphasize
that once $J'$ is introduced, however, the string-order parameter should
vanish even in the H-I phase, being similar to the $S=1$
ladder.\cite{ladder-paper} Still one may consider that the two phases
essentially differ with each other since they are separated by the
AF-LRO phase. If, however, we introduce the bond alternation also in the
$y$-direction, the two phases can be connected without
passing the gapless state as explained below.

\begin{figure}[t]
 \hspace*{1em}(a) \hspace*{10.5em}(b)

 \vspace*{-.5em}
 \centerline{
 \epsfysize=5.7cm\epsfbox{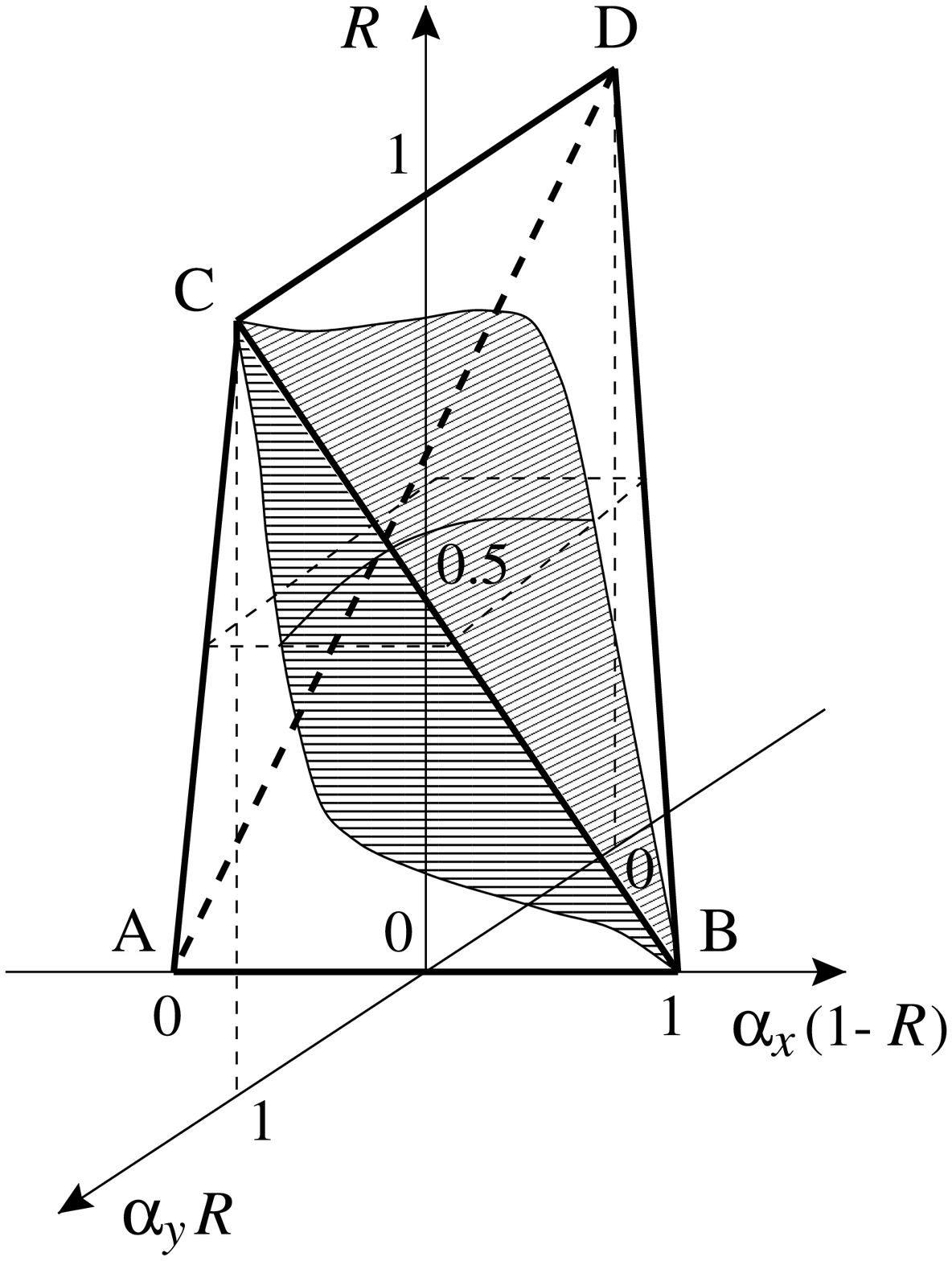}
 \epsfysize=5.7cm\epsfbox{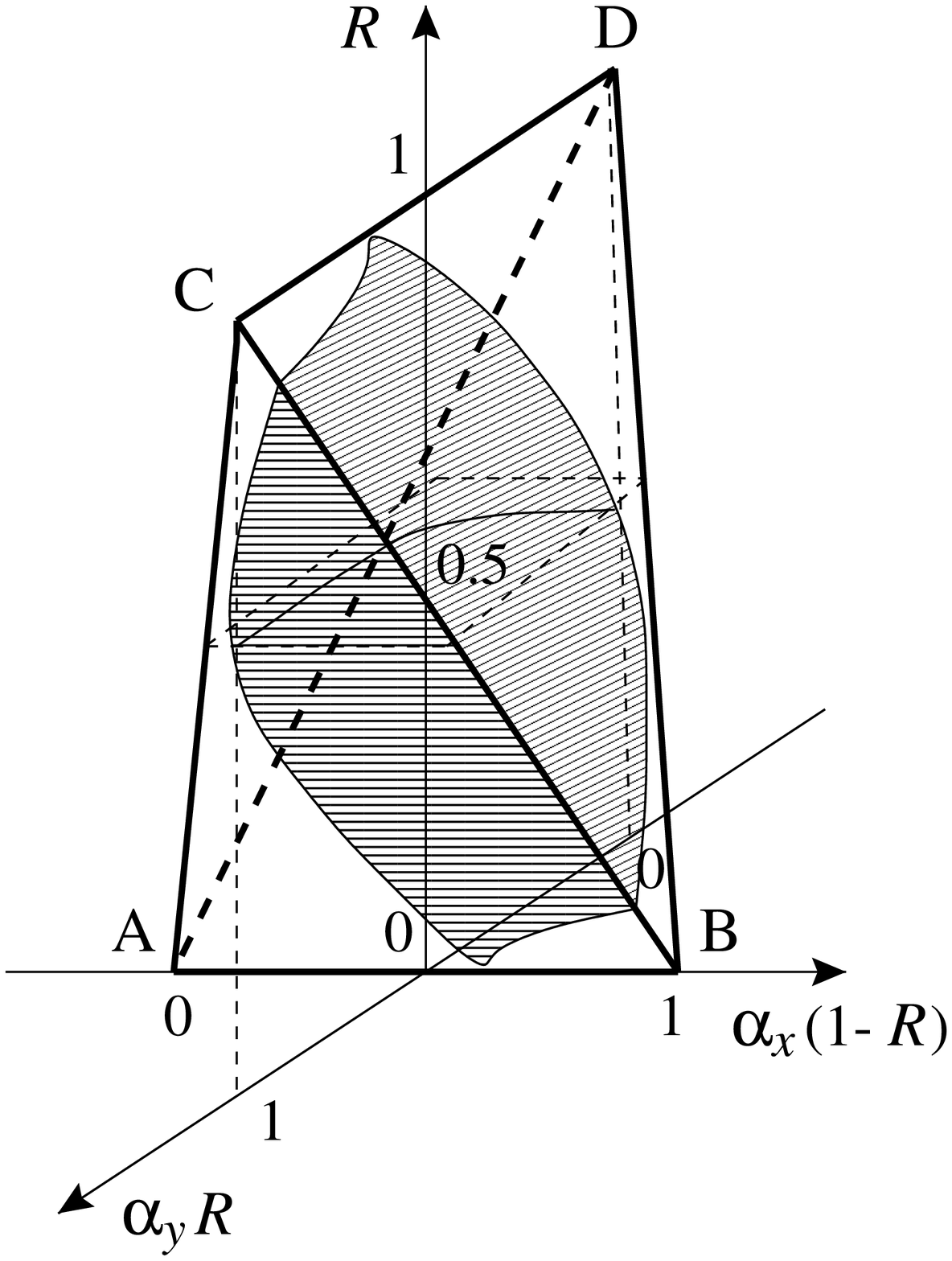}}
 \vspace*{.5em}
 \caption{Schematic ground-state phase diagram in the extended parameter
 space for (a) $S=1/2$ and (b) $S=1$.  The shaded areas denotes the
 AF-LRO phase on the ABC and BCD planes.  The cross section on the plane
 $R=0.5$ ($x$-$y$ isotropic plane) is drawn based on the numerical
 results.}
 \label{tetrapack_phase_diagram}
\end{figure}

Let us consider the 2D HAF model defined in the extended parameter space:
\begin{eqnarray}
\label{ham2}
{\cal H}&=&J_x\left\{\sum_{i,j}{\bf S}_{2i,j}\cdot{\bf S}_{2i+1,j}
+\alpha_x\sum_{i,j}{\bf S}_{2i+1,j}\cdot{\bf S}_{2i+2,j}\right\} \nonumber \\
&+&J_y\left\{\sum_{i,j}{\bf S}_{i,2j}\cdot{\bf S}_{i,2j+1}
+\alpha_y\sum_{i,j}{\bf S}_{i,2j+1}\cdot{\bf S}_{i,2j+2}\right\} \ .
\nonumber \\
\end{eqnarray}
The original Hamiltonian (\ref{ham}) corresponds to the case with
$J_x=1$, $\alpha_x = \alpha$, $J_y = J'$, and $\alpha_y=1$.  The
Hamiltonian~(\ref{ham2}) is invariant under the exchange between
$(J_x,\alpha_x)$ and $(J_y,\alpha_y)$.

To draw the phase diagram in this extended parameter space, it is convenient
to introduce a parameter, $R \equiv J_y/(J_x+J_y)$.  The limits
$J_y \rightarrow 0$ and $J_y \rightarrow \infty$ correspond to $R=0$
and 1, respectively.  Since $\alpha_x$ ($\alpha_y$) becomes irrelevant
in the limit $J_x \rightarrow 0$ ($J_x \rightarrow \infty$), the whole
phase diagram in the three-dimensional parameter space is shaped as a
tetrahedron.  In Fig.~\ref{tetrapack_phase_diagram}, we present the
ground-state phase diagram parametrized by $R$, $\alpha_x (1-R)$, and
$\alpha_y R$ for $S=1/2$ (a) and $S=1$ (b).  It should be noted that the
phase diagram should be invariant under the transformation, $(R,\alpha_x
(1-R),\alpha_y R) \leftrightarrow (1-R,\alpha_y R,\alpha_x (1-R))$,
reflecting the symmetry in the Hamiltonian explained above.  In the
phase diagram, the edge AB (CD) corresponds to isolated $x$-parallel
($y$-parallel) decoupled chains, the edge AD isolated four-spin
plaquettes, and the edge AC (BD) the two-leg ladders in $y$-direction
($x$-direction).

The face ABC (and also CDB) in Fig.~\ref{tetrapack_phase_diagram}
corresponds to the original phase diagram shown in
Figs.~\ref{gs_phase_diagram_spin_half} and
\ref{gs_phase_diagram_spin_one}, though the $y$-parallel-chain limit
$J'\rightarrow\infty$ in the original diagram is represented by one
vertex C (B) in the new ones.  In the extended phase diagram the shaded
(unshaded) area represents the AF-LRO (spin-gapped) phase on the ABC-
and CDB-faces.

It should be emphasized that on the ACD- and DBA-faces there is no
AF-LRO phase, since the system is one-dimensional dimerized two-leg
ladder.  There exist only the 1D critical points discussed
already.  Especially, there is no critical point on the edge AD.
Therefore, in the $S=1$ case, the three spin-gapped phases, H-I, D, and
H-II, are connected by the path
C$\rightarrow$A$\rightarrow$D$\rightarrow$B.  Similarly, in the $S=1/2$
case, the two dimer phases, which correspond to the vertex A and D,
respectively, are connected directly by the path A$\rightarrow$D.  Thus,
in both cases, there are only two phases, namely, the spin-gapped phase
and the AF-LRO one.

\section{Concluding Remarks}
\label{conclusion-section}

In this paper, we have investigated the ground-state phase diagram of
$S=1/2$ and $S=1$ HAF on the anisotropic dimerized square lattice by
means of the extensive QMC simulation with the continuous-imaginary-time
loop algorithm and the FSS analyses. It is confirmed that, for both
$S=$1/2 and 1, the quantum critical phenomena in the model belong to the
same universality class as that of the 3D classical Heisenberg model,
except for the 1D critical points, which belong to the same universality
class as that of the Gaussian model.  We have also demonstrated that the
spin-gapped phases of the 1D chain are connected when we introduce the
interchain couplings with bond alternation.  In the 2D system, only one
spin-gapped phase exists in both of the $S=1/2$ and $S=1$ systems.

The results obtained in the present work are considered to be the proper
basis for investigation of peculiar phenomena observed in the Q1D HAF's
mentioned at the beginning of this paper. We have already reported the
QMC analysis on the site-dilution-induced AF LRO in these materials
based on the Hamiltonian (\ref{ham}).~\cite{yasuda} In order to discuss
various experimental results quantitatively, it is certainly necessary
to take into account other ingredients than in Eq.~(\ref{ham}), such as the
next-nearest-neighbor intrachain interaction and the single-ion
anisotropy. They are beyond the scope of the present work. The present
results, however, demonstrate the role of the higher dimensionality
which has been overlooked so far.

Most of the numerical calculations in the present work have been
performed on the DEC Alpha, SGI ORIGIN 2000, SGI 2800, and RANDOM at the
Materials Design and Characterization Laboratory, Institute for Solid
State Physics, University of Tokyo and on the Hitachi SR-2201 at the
Supercomputer Center, University of Tokyo. The program used in the
present simulation was based on the library `Looper version 2' developed
by S.T. and K. Kato and also on the `PARAPACK version 2' by S.T.  The
present work was supported by the ``Research for the Future Program''
(JSPS-RFTF97P01103) of the Japan Society for the Promotion of Science.
S.T's work was partly supported by the Swiss National Science
Foundation.

\end{document}